\documentclass[conference]{IEEEtran}
\IEEEoverridecommandlockouts
\usepackage{cite}
\usepackage{amsmath,amssymb,amsfonts}
\usepackage{algorithm}
\usepackage{algpseudocode}
\usepackage{graphicx}
\usepackage{textcomp}
\usepackage{xcolor}
\usepackage{tikz}
\usetikzlibrary{shapes.geometric, arrows.meta, positioning}
\usepackage{subcaption}  

\usepackage{tikz}
\usepackage[T1]{fontenc}
\usepackage{tikz}
\usetikzlibrary{shapes.geometric, arrows.meta, positioning}
\usetikzlibrary{decorations.markings} 

\tikzset{
	startstop/.style={
		ellipse, draw, top color=red!10, bottom color=red!40,
		minimum width=2.5cm, minimum height=0.85cm,
		font=\footnotesize, align=center
	},
	fcblock/.style={
		rectangle, rounded corners=6pt,
		draw, top color=white, bottom color=blue!15!gray!30,
		minimum width=2.7cm, minimum height=0.85cm,
		font=\footnotesize, align=center
	},
	fcdec/.style={
		diamond, aspect=2, draw,
		top color=green!10, bottom color=brown!30,
		minimum height=1.15cm, font=\footnotesize, align=center
	},
	fcarrow/.style={draw=black!75, -{Latex[length=3mm, width=2mm]},  thick},
}

\def\BibTeX{{\rm B\kern-.05em{\sc i\kern-.025em b}\kern-.08em
		T\kern-.1667em\lower.7ex\hbox{E}\kern-.125emX}}
\begin{document}
	\title{
		Max-Min Fairness-Oriented Beamforming Design in HAPS-Enabled ISAC for 6G Networks
	}
\author{
	\IEEEauthorblockN{
		Parisa Kanani\IEEEauthorrefmark{1},
		Mohammad Javad Omidi\IEEEauthorrefmark{1}\IEEEauthorrefmark{2},
		Mahmoud Modarres-Hashemi\IEEEauthorrefmark{1}, and
		Halim Yanikomeroglu\IEEEauthorrefmark{3}
	}
	\IEEEauthorblockA{\footnotesize 
		\IEEEauthorrefmark{1}\textit{Department of Electrical and Computer Engineering, Isfahan University of Technology, Isfahan, Iran} \\
		\IEEEauthorrefmark{2}\textit{Department of Electronics and Communication Engineering, Kuwait College of Science and Technology, Doha, Kuwait} \\
		\IEEEauthorrefmark{3}\textit{Non-Terrestrial Networks (NTN) Lab, Department of Systems and Computer Engineering, Carleton University, Ottawa, Canada}
	}
}

		\maketitle
	\begin{abstract}
		This paper presents a high-altitude platform station (HAPS)-enabled integrated sensing and communication (ISAC) system designed for sixth-generation (6G) networks. Positioned in the stratosphere, HAPS serves  as a super-macro base station, leveraging advanced beamforming techniques to enable communication and sensing simultaneously.
		This research addresses the need for equitable service distribution in 6G networks by focusing on fairness within the HAPS-ISAC system. It tackles a non-convex optimization problem that balances  sensing beampattern gain and signal-to-interference-plus-noise ratio (SINR) requirements among communication users (CUs) using a max-min fairness approach while adhering to power constraints.
		The proposed HAPS-ISAC framework ensures efficient resource allocation, reliable coverage, and improved sensing accuracy.
		Simulation results validate the potential of HAPS-ISAC as a pivotal enabler for 6G networks and integrated communication-sensing systems.
			\end{abstract}
		
		\section{Introduction} 
The sixth-generation (6G)  wireless technology aims to achieve significantly higher data rates and expand coverage areas. Non-terrestrial networks (NTN) are expected to play a crucial role in 6G by extending coverage, enhancing connectivity, and addressing capacity demands. Among these, high-altitude platform station (HAPS) systems have emerged as a promising technology, positioned at altitudes between 20 and 50 kilometers. These platforms offer distinct advantages over traditional satellite and terrestrial networks, benefiting from better communication conditions and stable, quasi-stationary positions, enabling them to deliver precise and efficient services \cite{2,3,222}.  Compared to satellites, HAPS systems offer lower latency, reduced signal transmission delay, and lower construction and maintenance costs \cite{3,22,23,24}. In comparison to terrestrial stations, HAPS systems enable easier deployment and more stable long-term coverage, making them particularly suitable for dense urban environments \cite{22,24}. Additionally, unlike uncrewed aerial vehicles (UAVs), HAPS systems  benefit from a continuous power supply and extended operational endurance, making them superior for large-scale, long-term missions \cite{222,hapsisac,25}. 
Combining HAPS with integrated sensing and communication (ISAC) can lead to enhanced spectral efficiency, improved signal quality, and reduced latency.

	ISAC is an emerging technology that combines sensing and communication systems into a unified framework, enabling them to share frequency bands and hardware resources. 
	This integration enhances energy efficiency, reduces hardware costs, and boosts spectral efficiency, supporting higher data rates and improved network performance. ISAC allows a single radio frequency signal to simultaneously transmit both sensing and communication data, optimizing resource utilization \cite{5,6,2}. It has gained significant attention from both industry and academia for its potential to revolutionize network architectures, particularly in advancing wireless networks from fifth-generation (5G) to 6G systems.  Additionally, ISAC supports precise localization, advanced beamforming, efficient channel state information (CSI) tracking, and environmental reconstruction, all of which contribute to improved communication performance~\cite{7,10,8}.
		
		Despite the growing interest in ISAC, most existing research focuses exclusively on terrestrial networks \cite{pr1,pr2,10023}. For instance, a terrestrial reconfigurable intelligent surface (RIS)-assisted multiple-input multiple-output (MIMO) ISAC system is studied in \cite{pr1}, where an iterative algorithm is proposed to maximize sensing mutual information under quality-of-service (QoS) and hardware constraints.
 Reference \cite{pr2} investigates a ground-based ISAC system using orthogonal frequency-division multiplexing (OFDM) waveforms and introduces a deep reinforcement learning strategy for adaptive resource allocation in radar sensing under tracking and signal-to-noise ratio (SNR) constraints. In \cite{10023}, a beamforming technique is developed to jointly minimize power consumption and maximize the communication sum rate while ensuring signal-to-interference-plus-noise ratio (SINR) requirements for both radar sensing and communication. However, terrestrial ISAC systems suffer from limited coverage and line-of-sight (LoS) availability, motivating increased attention to NTN-based solutions, particularly using satellites and UAVs \cite{20,5003,pr3}.
 For instance, \cite{5003} addresses interference management in quantized ISAC-low Earth orbit (LEO) systems by employing rate-splitting multiple access (RSMA) to improve energy efficiency and sensing performance under power constraints. Nonetheless, ensuring reliable sensing and efficient power usage over long LEO satellite links remains challenging.
In \cite{20}, a joint UAV trajectory and beamforming optimization approach is proposed to maximize throughput while preserving radar beampattern gains using unified ISAC signals. Reference \cite{pr3} focuses on a multi-antenna UAV-enabled ISAC system, optimizing both communication/sensing precoding and flight trajectory to simultaneously maximize the minimum data rate among communication users and the minimum target detection probability, considering UAV energy limitations. 
However, due to battery constraints, UAV-based ISAC systems face scalability and endurance limitations.
 Integrating ISAC with NTNs introduces additional challenges in equitable resource allocation, particularly in balancing sensing beampattern gain and communication SINR for 6G networks.
  HAPS systems, with their stratospheric positioning, wide coverage, and extended endurance, offer a promising alternative for fairness-driven ISAC frameworks, enhancing sensing precision and communication reliability.
  However, the potential of HAPS for fairness-oriented ISAC remains largely unexplored, highlighting an open research direction for future 6G networks.

			This paper proposes a HAPS-enabled ISAC framework based on a multiple-input single-output (MISO)  architecture, where HAPS acts as a macro base station. The proposed system simultaneously supports downlink communication with single-antenna communication
			users (CUs) and radar target sensing, utilizing beamforming with  uniform planar array (UPA) steering vectors to ensure signal alignment. The primary objective is to maximize the minimum beampattern gain for sensing coverage while satisfying SINR requirements for CUs and adhering to the power limitations of the HAPS. This approach aims to improve sensing accuracy and communication reliability, fully leveraging the potential of HAPS-enabled ISAC systems for future 6G networks.
			
			The structure of the paper is organized as follows: Section II details the HAPS-MISO-enabled ISAC system model. Section III formulates the optimization problem aimed at maximizing the minimum beampattern gain while adhering to SINR and power constraints. Section IV discusses the simulation methodology and evaluates the results, demonstrating the effectiveness of the proposed scheme. Finally, conclusions are drawn in Section \ref{con} based on the presented findings.
			\section{System Model}
			The system model integrates HAPS within an ISAC framework, where the HAPS acts as a super macro base station. It supports communication for \( K \) single-antenna users while simultaneously performing radar sensing to detect \( J \) ground targets. This dual functionality enhances resource efficiency, allowing communication and sensing to occur together. The HAPS is equipped with a UPA consisting of \( S = S_W \times S_L \) antenna elements, where \( S_W \) and \( S_L \) denote the array’s width and length, respectively.
			
			The antenna array on the HAPS has element spacing \( d_l = d_w = \frac{\lambda}{2} = d \), where \( \lambda = \frac{c}{f} \) is the wavelength and \( c = 3 \times 10^8 \, \text{m/s} \) is the speed of light. The CUs are indexed by \( k \in \mathbb{K} = \{1, 2, \dots, K\} \), and potential targets are indexed by \( j \in \mathbb{J} = \{1, 2, \dots, J\} \).
			
			The system utilizes a Rician channel model that incorporates both line-of-sight (LoS) and non-line-of-sight (NLoS) components \cite{27}, capturing all relevant signal paths influencing communication. During time slot $n$, the HAPS transmits signals for communication and sensing, expressed as $\mathbf{x}[n] = \sum_{k=1}^{K} \mathbf{w}_k [n] s_k [n] + \sum_{j=1}^{J} \mathbf{r}_j [n] s'_j [n], \quad \forall n \in \mathbb{N}$.
Here, $\mathbf{w}_k[n] \in \mathbb{C}^{S \times 1}$ and $\mathbf{r}_j[n] \in \mathbb{C}^{S \times 1}$ denote the beamforming vectors for communication with user $k$ and radar sensing for target $j$, respectively. The signals $s_k[n]$ and $s'_j[n]$, representing the communication signal for user $k$ and the radar signal for target $j$, respectively, are modeled as independent random variables with zero mean and unit variance.
			
			The received signal at CU $k$ during time slot $n$ is given by $z_k[n] = \mathbf{h}_k^H [n] \mathbf{x}[n] + v_k[n]$, where $v_k[n]$ denotes 
			additive white Gaussian noise
			(AWGN) with zero mean and variance $\sigma_k^2$ at the receiver.
			
			The channel vector between the HAPS and the $k$-th CU is given by   \cite{29,30}:
			\begin{equation}
				\mathbf{h}_k = \frac{1}{\sqrt{\partial_k}} \left( \sqrt{\frac{K_k}{1+K_k}}\,\mathbf{h}_{k,\mathrm{LoS}} + \sqrt{\frac{1}{1+K_k}}\,\mathbf{h}_{k,\mathrm{NLoS}} \right),
			\end{equation}
			where $\partial_k = \left(\frac{4\pi r_k}{\lambda}\right)^2$ represents free-space path loss (FSPL), $K_k$ is the Rician factor, and $r_k$ is the distance between the HAPS and the CU $k$. The LoS component $\mathbf{h}_{k,\mathrm{LoS}}$ is deterministic, while the NLoS component $\mathbf{h}_{k,\mathrm{NLoS}}$ consists of elements that are distributed  as a complex Gaussian random variable with zero mean and unit variance.
			The LoS component $\mathbf{h}_{k,\mathrm{LoS}}$ is expressed as  
			\begin{equation}
				\mathbf{h}_{k,\mathrm{LoS}} = \boldsymbol{\alpha}(\theta_k, \phi_k) \otimes \mathbf{b}(\theta_k, \phi_k),
				\label{h}
			\end{equation}
			where  
			\begin{equation}
				\boldsymbol{\alpha} (\theta_k, \phi_k) = 
				\begin{bmatrix}
					1, \ e^{-j2\pi(d\sin\theta_k\cos\phi_k)/\lambda}, \ \cdots
				\end{bmatrix}^T,
			\end{equation}
			and  
			\begin{equation}
				\mathbf{b} (\theta_k, \phi_k) = 
				\begin{bmatrix}
					1, \ e^{-j2\pi(d\sin\theta_k\sin\phi_k)/\lambda}, \ \cdots
				\end{bmatrix}^T.
			\end{equation}  
			Here, $\theta_k$ and $\phi_k$ represent the vertical and horizontal angles of departure (AoD) of the $k$-th CU, respectively \cite{29,30}. The Kronecker product $\otimes$ combines the vectors.

			The SINR at CU $k$ is given by  
			\begin{align}
			\text{SINR}_k = \gamma_k[n] = \frac{\left| \mathbf{h}_k^H[n] \mathbf{w}_k[n] \right|^2}{I_k[n] + \sigma_k^2},
			\end{align}
			where  
			\begin{align*}
				I_k[n] &=  \sum_{i=1, i \neq k}^{K} \left|\mathbf{h}_k^H[n] \mathbf{w}_i[n] \right|^2+ \sum_{j=1}^{J}\left| \mathbf{h}_k^H[n] \mathbf{r}_j[n] \right|^2.
			\end{align*}
			The achievable rate at CU $k$ in time slot $n$ is then given by $R_k[n] = \log_2(1 + \gamma_k[n])$.
			
			The average transmitted power from the HAPS during time slot $n$, constrained by $P_{\text{max}}$, is $ \mathbb{E}(\|\mathbf{x}[n]\|^2) = \sum_{k=1}^{K} \|\mathbf{w}_k[n]\|^2 + \sum_{j=1}^{J} \|\mathbf{r}_j[n]\|^2 \le P_{\text{max}}, \quad \forall n \in \mathbb{N}. $
			This ensures the total power used for communication and sensing does not exceed the maximum limit.
			
			In the sensing phase, HAPS operates using a cellular approach to detect potential targets at $J$ predefined locations on the ground, denoted by $\mathbf{m}_j$ for $j \in \{1, \ldots, J\}$, according to specific sensing tasks.
			To improve sensing coverage, the goal is to maximize the minimum beampattern gain towards each $\mathbf{m}_j$. The beampattern gain  is given by \cite{7,20}
			\begin{align} 
				\zeta([n], \mathbf{m}_j) &= \mathbb{E}\left[ \left| \mathbf{a}^H([n], \mathbf{m}_j) \mathbf{x}[n] \right|^2 \right]\nonumber\\
				& = \mathbf{a}^H([n], \mathbf{m}_j) \bigg(   \sum_{k=1}^{K}\mathbf{w}_k [n]\mathbf{w}^H_k [n] \nonumber \\ & + \sum_{j=1}^{J}\mathbf{r}_j [n]\mathbf{r}^H_j [n]  \bigg) \mathbf{a}([n], \mathbf{m}_j),
				\label{beam}
			\end{align}
			where $\mathbf{a}([n], \mathbf{m}_j)$ denotes the steering vector from HAPS to $\mathbf{m}_j$ at time $n$, as defined in \eqref{h}.
				\section{Problem Formulation}
			\label{problem}
			The main objective of this study is to maximize the minimum beampattern gain for the target while satisfying the SINR requirements for CUs. This is formulated as the following optimization problem
			\begin{align}
				\max_{\underset{ k \in \mathbb{K},j \in \mathbb{J}}{\mathrm{\mathbf{w}_k [n], \mathbf{r}_j [n]}}} & \: \min_j \quad  \zeta_j([n], \mathbf{m}_j) \label{mfi} \\
				\text{s.t.} \quad\quad\, & \: \sum_{k=1}^{K} \| \mathbf{w}_k[n] \|^2 + \sum_{j=1}^{J} \| \mathbf{r}_j [n] \|^2  \leq P_{\max}, \nonumber\\ 
				& \: \forall  n \in \mathbb{N}, j \in \mathbb{J}, k \in \mathbb{K},  \tag{a}\\
				& \: \text{SINR}_\text{th} \leq \text{SINR}_k	 \quad \forall k \in \mathbb{K} \tag{b}.
			\end{align}
			The transmitted power constraint in (\ref{mfi}.a) limits the maximum power, \(P_{\max}\), that HAPS can transmit, ensuring safe and acceptable operating levels. Constraint (\ref{mfi}.b) must be met for all CUs, with \(\text{SINR}_{\text{th}}\) representing the predefined SINR threshold. 
			The optimization problem in \eqref{mfi} is non-convex and NP-hard due to the coupled design of sensing and communication beamformers under a max-min fairness criterion \cite{boyd}.
			 This coupling creates a highly complex search space whose size grows exponentially with the number of CUs $K$, targets $J$, and antennas $S$, making exact solutions computationally intractable.

			To improve the tractability of the optimization, we introduce an auxiliary variable \(\eta\) to capture the minimum beampattern gain, reformulating the problem to maximize \(\eta\) for a more efficient solution:
			\begin{align}
				\max_{\underset{ k \in \mathbb{K},j \in \mathbb{J}}{\mathrm{\mathbf{w}_k [n], \mathbf{r}_j [n], \eta}}} & \quad \quad \eta \ \label{mfi2} \\
				\text{s.t.} \quad\quad\quad\quad \, &\eta \leq \zeta_j(\mathbf{q}[n], \mathbf{m}_j) \quad \forall j \in \mathbb{J}, n \in \mathbb{N},  \tag{a}\\
				\quad\quad\quad\quad \, &\sum_{k=1}^{K} \| \mathbf{w}_k[n] \|^2 + \sum_{j=1}^{J} \| \mathbf{r}_j [n] \|^2  \leq P_{\max}, \nonumber\\& \forall n \in \mathbb{N}, j \in \mathbb{J}, k \in \mathbb{K},  \tag{b}\\
				&  \text{SINR}_\text{th} \leq \text{SINR}_k \quad \forall k \in \mathbb{K}, \tag{c}
			\end{align}
%
%
%
			Here, \(\eta\) serves as an auxiliary variable representing the minimum beampattern gain across all targets, transforming the original max-min problem into a maximization framework \cite{boyd}. This reformulation provides a more structured problem representation that facilitates the design of efficient solution strategies. However, the problem remains inherently non-convex and computationally challenging due to the joint optimization of communication and sensing beamformers, along with the nonlinear SINR constraints expressed as fractional quadratic functions. Consequently, despite the improved tractability, obtaining globally optimal solutions remains difficult in practice.

				\section{Performance Assessment}
				In this section, we present simulation results to evaluate the performance of the proposed HAPS-enabled ISAC system. 
				The simulation involves randomly placed CUs and targets within a 1 km² network area, with parameters specified in Table \ref{combined} unless otherwise noted. The HAPS is assumed to be centrally located relative to all service areas.
				The goal is to optimize key variables, such as the transmit beamforming vectors \(\mathbf{w}_k[n]\) and \(\mathbf{r}_j[n]\) to achieve optimal system performance.

				\subsection{Optimizing HAPS-ISAC  Network Using Genetic Algorithm}

				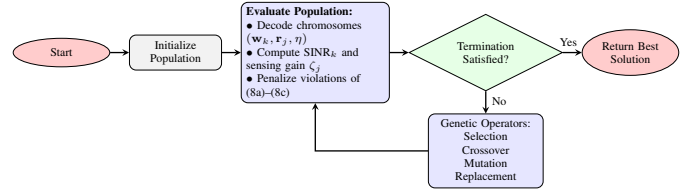
\begin{figure}[t]
					\centering
					\resizebox{\columnwidth}{!}{%
						\begin{tikzpicture}[node distance=1cm and 0.4cm,
							every node/.style={font=\scriptsize},
							multi_block/.style={
								rectangle, rounded corners, draw, fill=blue!10,
								text width=2.9cm, align=left, inner sep=3pt, minimum height=1.6cm
							},
							op_block/.style={
								rectangle, rounded corners, draw, fill=blue!10,
								text width=2.2cm, align=center, inner sep=3pt, minimum height=1.2cm
							},
							fcblock/.style={
								rectangle, rounded corners, draw, fill=gray!10,
								align=center, minimum height=0.6cm, text width=1.7cm
							},
							fcdec/.style={
								diamond, aspect=2.0, draw, fill=green!10, align=center,
								inner sep=2pt, text width=1.9cm
							},
							startstop/.style={
								ellipse, draw, fill=red!20, align=center,
								text width=1.2cm, minimum height=0.6cm
							},
							fcarrow/.style={->,>=stealth,thick}]
							
							\node[startstop] (start) {Start};
							\node[fcblock, right=of start] (init) {Initialize\\Population};
							\node[multi_block, right=of init] (eval) {
								\textbf{Evaluate Population:}\\
								$\bullet$ Decode chromosomes $(\mathbf{w}_k,\mathbf{r}_j,\eta)$\\
								$\bullet$ Compute SINR$_k$ and sensing gain $\zeta_j$\\
								$\bullet$ Penalize violations of (8a)--(8c)
							};
							\node[fcdec, right=of eval] (term) {Termination\\Satisfied?};
							\node[startstop, right=of term] (end) {Return Best\\Solution};

							\node[op_block, below=0.47cm of term] (ops) { 
								Genetic Operators:\\
								Selection\\Crossover\\Mutation\\Replacement
							};
							
							\draw[fcarrow] (start)--(init);
							\draw[fcarrow] (init)--(eval);
							\draw[fcarrow] (eval)--(term);
							\draw[fcarrow] (term)--node[above,pos=0.3]{Yes}(end);
							\draw[fcarrow] (term.south) -- ++(0,-0.21cm) node[right]{No} -| (ops.north);
							\draw[fcarrow] (ops.west)-|(eval.south);
							
					\end{tikzpicture}}
					\caption{GA flow for max--min beampattern gain optimization
					under	performance constraints.}
					\label{fig:ga-flow-final}
				\end{figure}
				\begin{algorithm}[t]
				\centering
				\footnotesize 
				\caption{GA for the Max--Min Fairness Problem}
				\label{alg:generic-ga}
				\begin{algorithmic}[1]
					\setlength{\itemsep}{0pt} 
					\State \textsc{Initialize} a population of candidate solutions
					\For{each generation (\(1\) to \( \text{Max\_Generations} \))}
					\If{termination criterion met (\textit{e.g.,~desired objective achieved})}
					\State \textbf{break}
					\EndIf
					\State \textsc{Evaluate} the quality (fitness) of each solution
					\State \textsc{Select} high-quality solutions for variation
					\State \textsc{Apply crossover} between selected solutions with a predefined probability
					\State \textsc{Apply mutation} to the resulting solutions with a mutation probability
					\State \textsc{Update} the population for the next generation
					\EndFor
					\State \textsc{Return} the best solution found
				\end{algorithmic}
					\end{algorithm}
%
The problem discussed in Section~\ref{problem} involves solving the non-convex optimization in \eqref{mfi2}, which is computationally intractable and NP-hard due to its complexity and high dimensionality \cite{boyd, ga33}. Traditional convex methods are often inadequate in such scenarios due to their limited global search capabilities.
To address this, we adopt a metaheuristic approach using the genetic algorithm (GA)—a robust, scalable tool effective in power allocation for both CUs and targets \cite{gahaps, ga33}. Its flexibility and ease of implementation further make it well-suited for tackling the non-convexity and high-dimensional nature of the problem.
The GA operates by iteratively evolving solutions based on the principles of natural selection, as outlined in Algorithm~\ref{alg:generic-ga}. Starting from a random population, it applies selection, crossover, and mutation to improve candidate solutions over generations. Thanks to its global search capability, the GA is particularly well-suited for complex tasks such as path planning and task allocation~\cite{gauuav, book_ga}. Moreover, it holds great potential for power allocation and joint optimization problems in ISAC systems.

					In this study, we employ a heuristic optimization approach to tackle the non-convexity of the problem. Specifically, we use MATLAB's \texttt{ga} function from the Optimization Toolbox, which iteratively refines candidate solutions through selection, crossover, and mutation. 
					This algorithm provides a flexible and efficient framework for finding optimal or near-optimal solutions in complex scenarios, as illustrated in Fig.~\ref{fig:ga-flow-final}.
					The configuration and parameters of the genetic algorithm are provided in Table~\ref{combined}. 
					As summarized in Table~\ref{combined}, the genetic algorithm is configured with a population size of 2500 and a maximum of 1500 generations to ensure thorough search of the solution space. A crossover fraction of 0.81 is used to maintain diversity, while Gaussian mutation with a standard deviation of 0.02 enables fine-grained exploration. The function tolerance is set to \(10^{-6}\) to support accurate convergence. These settings were empirically determined through iterative tests to strike a balance between performance and efficiency.

				\subsubsection{Computational Complexity Analysis}
			The computational complexity of the genetic algorithm primarily depends on three main factors: the population size $N$, the number of generations $G$, and the computational cost of evaluating the fitness function, denoted by $C$. Accordingly, the overall complexity of the algorithm can be approximated as $\mathcal{O}(N \times G \times C)$~\cite{pich1,pich2}.

					\subsection{Simulation Results and Performance Analysis}
				This subsection presents a detailed analysis of the simulation and experimental results, focusing on the performance evaluation of the proposed HAPS-enabled ISAC system.

				\begin{table}[b]
					\centering
					\caption{Simulation Parameters and Algorithm Settings}
					\label{combined}
					\begin{tabular}{|c|c|}
						\hline
						\textbf{Parameter} & \textbf{Value} \\
						\hline \hline
						Number of CUs ($K$) & 4 \\
						\hline
						Number of potential target points ($J$) & 4 \\
						\hline
						Antennas of HAPS along the width ($S_w$) & 8 \\
						\hline
						Antennas of HAPS along the length ($S_l$) & 8 \\
						\hline
						Maximum  power of HAPS ($P_{\max}$) & 52 dBm \\
						\hline
						Noise power at each CU receiver  & -110 dBm \cite{20} \\
						\hline
						Antenna spacing ($d$) & $\lambda/2$ \\
						\hline
						Carrier frequency ($f$) & $2.545 \times 10^9$ \cite{enh} \\
						\hline
						Flight altitude of HAPS ($H_{\text{HAPS}}$) & 20000 m \\
						\hline
						Rician factor ($K_k$) & 10 \cite{enh} \\
						\hline
						Function tolerance & $10^{-6}$ \\
						\hline
						Number of population & 2500 \\
						\hline
						Crossover fraction & 0.81 \\
						\hline
						Mutation function & Gaussian Mutation \\
						\hline
						Standard deviation of mutation & 0.02 \\
						\hline
						Generations & 1500 \\
						\hline
					\end{tabular}
				\end{table}
				
					\begin{figure}[t]
					\centering
					\includegraphics[width=0.47\textwidth]
					{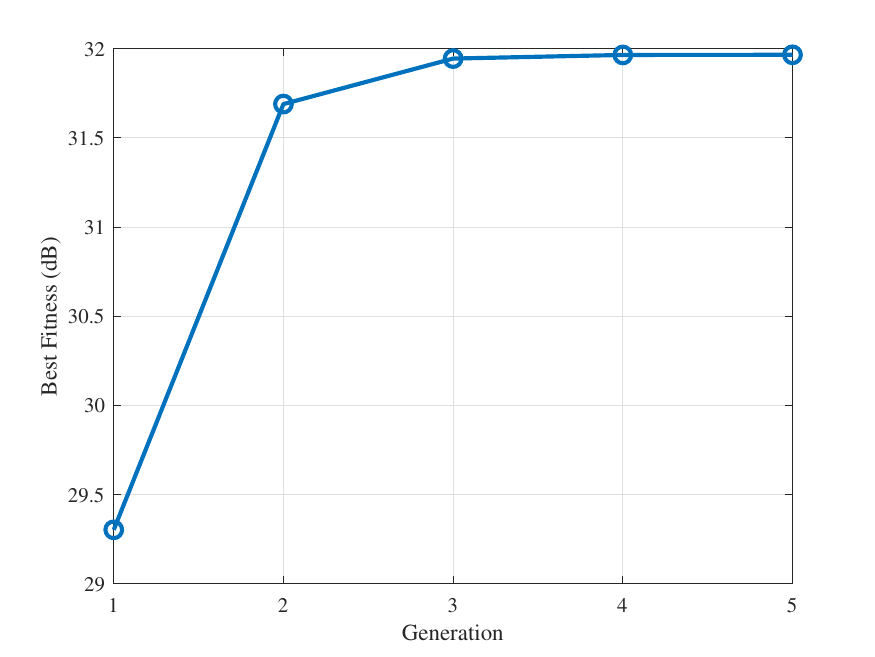}
				\caption{Convergence  of the genetic algorithm (GA) depicting the best objective function value in decibels (dB) over generations for the HAPS-enabled ISAC system.}
					\label{converg}
				\end{figure}
				
				\begin{figure}[t] 
					\centering
					\includegraphics[width=0.47\textwidth]{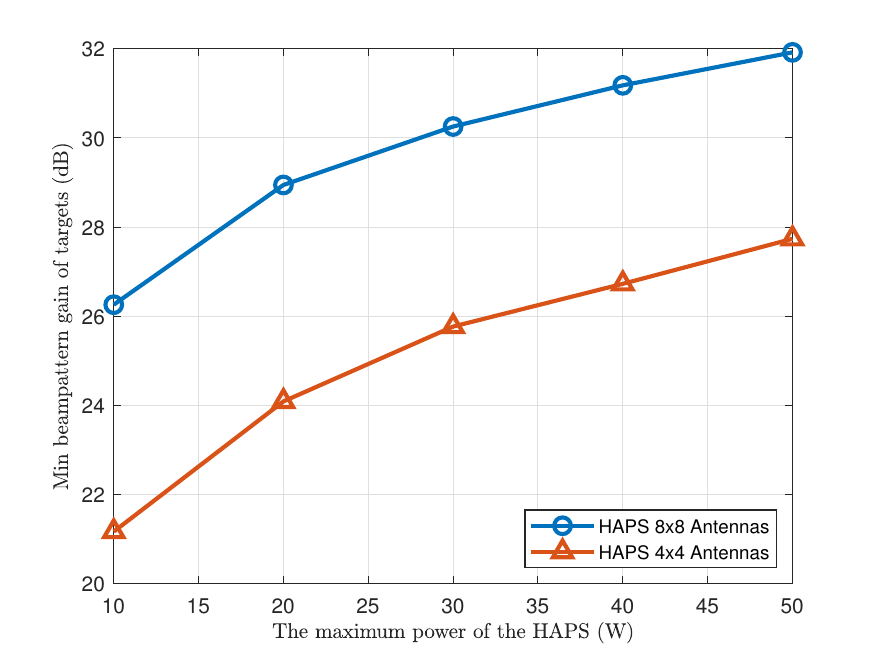}
					\caption{Variation of the objective function with power level for different antenna configurations.}
					\label{p_fig} 
				\end{figure}
				\begin{figure}[t]
					\centering
					\includegraphics[width=0.48\textwidth]{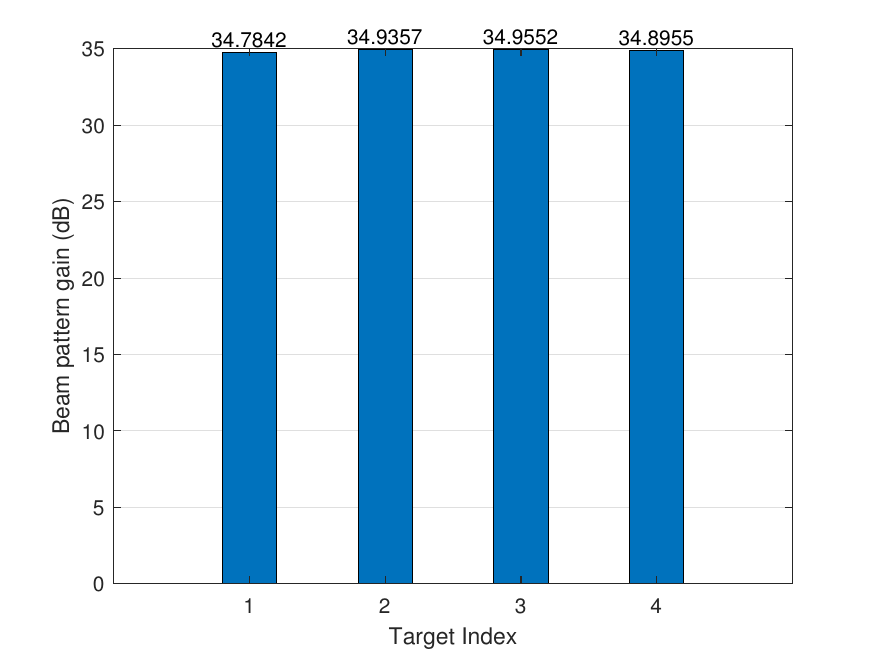}
					\caption{Fairness in beampattern gain towards desired sensing angles in the MAX-MIN optimization.}
					\label{fig:beam_pattern_gain}
				\end{figure}
				\begin{figure}[t]
					\centering
					\includegraphics[width=0.49\textwidth]{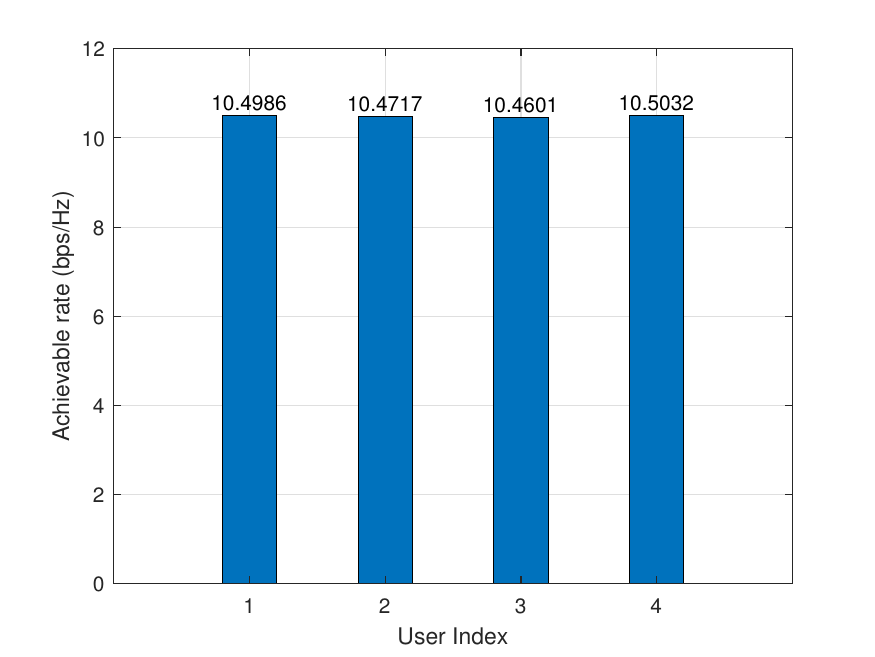}
					\caption{User rate distribution in the MAX-MIN beampattern gain optimization, showing fairness among CUs with optimized beamforming.}
					
					\label{fig:user_rate_distribution}
				\end{figure}
%
				\begin{figure}[t]
					\centering
					\includegraphics[width=0.47\textwidth]{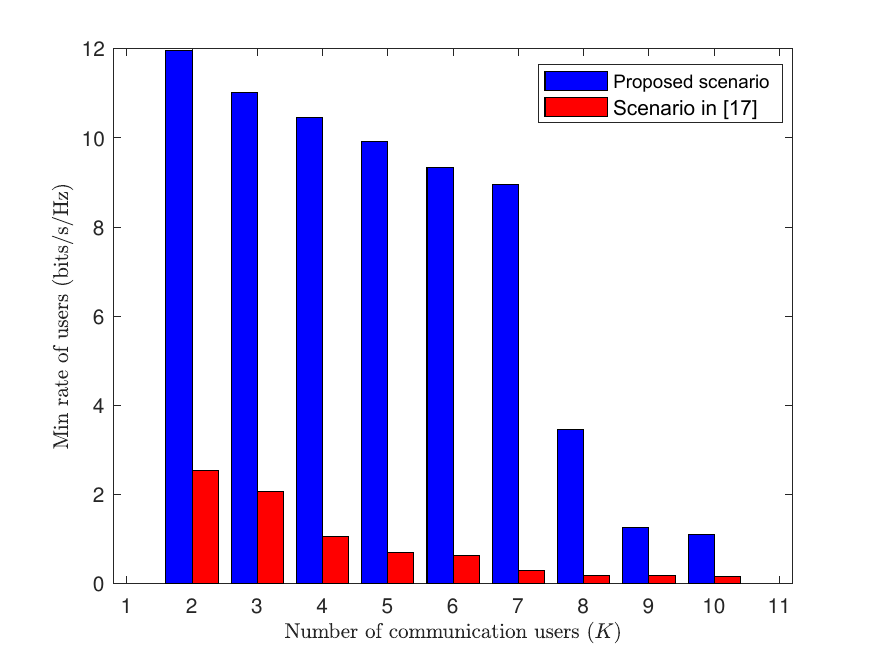}
					
					\caption{Comparison of the rate performance of the proposed model with that of \cite{20} based on the number of CUs 
						$K$.}
					\label{fig2_2}
				\end{figure}
				Fig.~\ref{converg} illustrates the convergence behavior of the GA employed to adress the $\max \eta$ optimization problem. 
				 The plot depicts the best-achieved value of $\eta$, expressed in decibels (dB), over five generations. Since the GA minimizes the objective function by default, $-\eta$ was used as the fitness function, and the output values were negated to reflect the true $\eta$ in decibels (dB).
				 This progression demonstrates the GA's effectiveness in optimizing the beamforming configuration to enhance the worst-case sensing performance metric  within the HAPS-ISAC system.
%
%
 The iterative improvement shown in the plot indicates the GA's suitability for tackling this complex optimization problem. Furthermore, the relatively rapid convergence within these generations suggests the GA's efficiency in attaining near-optimal beamforming solutions, a crucial aspect for potential real-time HAPS-ISAC applications.
				
				Fig. \ref{p_fig} shows that performance improves with higher power levels. Moreover, a higher number of antennas (8x8 and 4x4 configurations) enhances performance, with the 8x8 configuration showing the most significant improvement.
				
The effectiveness of the proposed HAPS-enabled ISAC framework in achieving fairness is clearly demonstrated in both \figurename~\ref{fig:beam_pattern_gain} and \figurename~\ref{fig:user_rate_distribution}. Specifically, \figurename~\ref{fig:beam_pattern_gain} illustrates how the optimization framework ensures balanced sensing performance by maximizing the minimum beampattern gain across all designated target angles. This guarantees uniform sensing quality while respecting key system constraints, such as the total transmit power and the SINR requirements of CUs. Simultaneously, \figurename~\ref{fig:user_rate_distribution} highlights the uniformity in communication rates among CUs, reflecting the robustness of the proposed strategy in maintaining service fairness even in resource-limited environments. The joint consideration of sensing and communication objectives within a unified beamforming design leads to an efficient allocation of spatial and power resources, ultimately enhancing  communication quality and system fairness  across varied scenarios.

	We compare our proposed approach with a UAV-based ISAC network that does not incorporate HAPS, which serves as a relevant baseline for evaluation given the current lack of extensive research on HAPS-assisted ISAC systems.
				Fig. \ref{fig2_2} compares the minimum user rates of our proposed method with those from Reference~\cite{20} across varying numbers of CUs ($K$). The genetic algorithm was applied to both models to ensure a consistent and fair comparison, demonstrating its capability to efficiently address complex, nonlinear optimization problems while enabling a uniform evaluation framework for performance assessment.
					In general, an increase in the number of CUs typically leads to a decline in the minimum user rate, particularly in resource-limited scenarios. As illustrated in the figure, this trend is observed in both models as the CU count rises. 
					However, our proposed model consistently outperforms the approach in Reference~\cite{20}, achieving a higher minimum user rate across all scenarios. This improvement underscores the enhanced fairness of our method. While this comparison focuses on the minimum user rate as an indicator of fairness, future studies could explore additional metrics, such as the sum rate, to provide a more comprehensive evaluation.

				\subsection{Practical Implementation Considerations}
					The GA involves significant computational demands and processing delays, challenging real-world applications. As optimization parameters such as the number of HAPS antennas, communication users, and sensing targets increase, complexity surges, requiring extended iterations and larger populations \cite{mobini}. Techniques including initial solution seeding, parameter tuning, distributed computing, hybrid algorithms, hardware acceleration, and dynamic termination criteria can alleviate these issues. When tailored specifically to max–min beampattern gain optimization, these strategies markedly improve computational efficiency, enabling effective and scalable HAPS-ISAC deployments \cite{warm1,warm2}.

				Furthermore, to provide a more comprehensive evaluation of the proposed approach, future work could benchmark its performance against fundamental theoretical limits. This would include deriving the Cram\'{e}r--Rao lower bound (CRLB) to quantify the lower bound on sensing accuracy, and establishing the communication capacity limits to characterize the achievable throughput under ideal conditions.

				\section{Conclusion}
				\label{con}

This paper proposed a HAPS-ISAC system tailored for 6G networks, in which beamforming was jointly optimized to enhance both communication and sensing performance. Simulation results demonstrated improvements in user throughput, beampattern gain, SINR, max-min fairness, and power allocation efficiency. Compared to conventional and UAV-based ISAC systems, the proposed approach exhibited superior scalability and more equitable resource distribution. These findings underscore the potential of HAPS-ISAC to meet the stringent requirements of 6G, enabling advanced applications through robust and fair connectivity. Future research can investigate the integration of reconfigurable intelligent surfaces, learning-based optimization strategies, and practical hardware constraints to enhance the performance and deployment feasibility of HAPS-ISAC systems in real-world 6G scenarios.

					\bibliographystyle{ieeetr}

				\bibliography{ref1_030808_2_main1_v2}
				
				\end{document}